\documentclass[tightenlines, floatfix, reprint, amssymb, amsmath, amsfonts, a4paper, aps, floatfix, superscriptaddress, showpacs, twoside, twocolumn, superscriptaddress, prx]{revtex4-1}

\usepackage{tabularx, xspace, rotating, units, enumerate, textcomp, icomma, hyperref}
\usepackage[T1]{fontenc}
\usepackage{acronym}
\usepackage[utf8]{inputenc}
\usepackage{amsmath, amssymb, amsfonts,amsbsy,amsthm}
\usepackage{graphicx,color}
\usepackage{float}
\usepackage{textgreek}

\definecolor{darkgreen}{rgb}{0.0, 0.5, 0.0}
\definecolor{purple}{rgb}{0.56, 0.0, 1.0}

\acrodef{QWs}{Quantum Wells}
\acrodef{QHE}{Quantum Hall Effect}
\acrodef{MBE}{Molecular Beam Epitaxy }
\acrodef{LLs}{Landau levels}

\newcommand{\IHPPAS}{\affiliation{Institute of High Pressure Physics, Polish Academy of Sciences, 01-142 Warsaw, Poland}}
\newcommand{\LCCNRS}{\affiliation{Laboratoire Charles Coulomb, University of Montpellier, 34950 Montpellier, France}}
\newcommand{\IPMRAS}{\affiliation{Institute for Physics of Microstructures, Russian Academy of Sciences, 603950 N. Novgorod,
Russia}}
\newcommand{\IPPAS}{\affiliation{Institute of Physics, Polish Academy of Sciences, PL02-668 Warsaw, Poland}}
\newcommand{\LUN}{\affiliation{Lobachevsky University, 603950 N. Novgorod, Russian Federation}}
\newcommand{\ISPPSB}{\affiliation{Institute of Semiconductor Physics, Russian Academy of Sciences, 630090 Novosibirsk, Russia}}
\newcommand{\TSURF}{\affiliation{Tomsk State University, 634050 Tomsk, Russia}}
\newcommand{\RISP}{\affiliation{Rzhanov Institute of Semiconductor Physics of the Siberian Branch of the RAS, 630090
Novosibirsk, Russia}}
\newcommand{\FMNSPL}{\affiliation{Faculty of Mathematics and Natural Sciences, Rzesz\'{o}w University, 35-959 Rzesz\'{o}w,
Poland}}
\newcommand{\WPIPL}{\affiliation{WPI-Advanced Institute for Materials Research, Tohoku University, Sendai 980-8577, Japan}}
\newcommand{\IRCPL}{\affiliation{International Research Centre MagTop, 02-668 Warsaw, Poland}}

\begin{document}

\title{Perspectives of HgTe Topological Insulators for Quantum Hall Metrology}

\author{Ivan Yahniuk}
\email{ivan.yahniuk@unipress.waw.pl}
\IHPPAS
\author{Sergey S. Krishtopenko}
\LCCNRS \IPMRAS
\author{Grzegorz Grabecki}
\IPPAS
\author{Benoit Jouault}
\LCCNRS
\author{Christophe Consejo}
\LCCNRS
\author{Wilfried Desrat}
\LCCNRS
\author{Magdalena Majewicz}
\IPPAS
\author{Alexander M. Kadykov}
\LCCNRS
\author{Kirill E. Spirin}
\IPMRAS
\author{Vladimir I. Gavrilenko}
\IPMRAS \LUN
\author{Nikolay N. Mikhailov}
\ISPPSB
\author{Sergey A. Dvoretsky}
\TSURF \RISP
\author{Dmytro B. But}
\IHPPAS
\author{Frederic Teppe}
\LCCNRS
\author{Jerzy Wr\'{o}bel}
\IPPAS \FMNSPL
\author{Grzegorz Cywi\'{n}ski}
\IHPPAS
\author{S\l{}awomir Kret}
\IPPAS
\author{Tomasz Dietl}
\IPPAS \WPIPL \IRCPL
\author{Wojciech Knap1}
\IHPPAS \LCCNRS
\email{knap.wojciech@gmail.com}

\begin{abstract}
We report the studies of high-quality HgTe/(Cd,Hg)Te quantum wells (QWs) with a width close to the critical one $d_c$, corresponding to the topological phase transition and graphene like band structure in view of their applications for Quantum Hall Effect (QHE) resistance standards. We show that in the case of inverted band ordering, the coexistence of conducting topological helical edge states together with QHE chiral states degrades the precision of the resistance quantization. By experimental and theoretical studies we demonstrate how one may reach very favorable conditions for the QHE resistance standards: low magnetic fields allowing to use permanent magnets (\emph{B} $\leq$ 1.4~T) and simultaneously realtively high teperatures (liquid helium, \emph{T} $\geq$ 1.3~K). This way we show that HgTe QW based QHE resistance standards may replace their graphene and GaAs counterparts and pave the way towards large scale fabrication and applications of QHE metrology devices.
\end{abstract}

\date{\today}

\maketitle

\section{INTRODUCTION} \label{intro}
Mercury cadmium telluride (Hg$_{1-x}$Cd$_{x}$Te) zinc-blende compounds are an example of rare semiconductor materials that form alloys over the whole composition range \emph{x} while keeping the same crystal structure and the virtually unaltered lattice parameters.~\cite{Dornhaus2006a, Galazka1967a} Accordingly, it is possible to tune the band structure by changing \emph{x} and grow bulk films, two-dimensional (2D) quantum wells (QWs) or superlattices without strain-related material degradation. In this sense, Hg$_{1-x}$Cd$_{x}$Te crystals are similar to the well-known Ga$_{1-x}$Al$_{x}$As semiconductors, but show a much larger
energy band-gap tunability, with band gaps ranging from E$_{g}$ $\equiv$ E$_{\Gamma6}$ - E$_{\Gamma8}$~= 1.6~eV for CdTe to the inverted band ordering, with E$_{g}$ $\approx$ -0.30~eV for HgTe at 4.2~ K.~\cite{Dornhaus2006a} This peculiar aspect of Hg$_{1-x}$Cd$_{x}$Te allows to reach  E$_g$ $\approx$ 0~eV~\cite{Galazka1967a, Orlita2014a, Teppe2016a} and the conditions for
observation of 3D carriers with massless Dirac-like linear dispersion and with high values of room- and low-temperature electron mobilities reaching 3.5$\cdot$10$^{4}$ and 2$\cdot$10$^{6}$~cm$^{2}$ V$^{-1}$s$^{-1}$, respectively~\cite{Dubowski1981a}. Moreover, since it is possible to adjust the bandgap below 100~meV, Hg$_{1-x}$Cd$_{x}$Te-based systems are broadly employed in infrared and terahertz detectors~\cite{Rogalski2016a}, cameras~\cite{Ruffenach2017a}, and lasers~\cite{Morozov2017a}.

Recent technological advances in molecular beam epitaxy (MBE) of Hg$_{1-x}$Cd$_{x}$Te-based quantum structures have opened new and striking possibilities. In particular, HgTe/(Cd,Hg)Te QWs have allowed the demonstration of the existence of various topological phases in condensed matter~\cite{Konig2007a, Konig2008a, Brune2011a}. By changing the QW widths, the barrier alloy composition and the number of coupled QWs, it has been possible to demonstrate 2D topological insulators with 1D edge conducting channels~\cite{Qi2011a, roche2015topological} as well as structures with band dispersions similar to single layer~\cite{Buttner2011a} or bilayer graphene~\cite{Krishtopenko2016b}.

Quantum Hall Effect (QHE) resistance standards make use of the accurate QHE quantization of the Hall resistance given by $\rho_{xy}$~= $\emph{h}/\emph{ie}^{2}$, where \emph{i} is an integer, and \emph{h} and \emph{e} are the Planck constant and the electron charge, respectively~\cite{Klitzing1980a,Tong2016a}. Since their discovery, QHE standards have become a highly important tool in quantum metrology and are widely used in many national/international standardization institutions today~\cite{Poirier2009a}. Currently, many materials are used for QHE standards. Despite differences in the material band structure, the sample imperfections and the geometry, unbelievably precise and wide Hall resistance plateaus are observed, independent of the host material, with a precision at the level going up to a few parts in 10~\cite{Konig2008a,
Jeckelmann2001a, Taylor1989a}.

\begin{figure}%[t!]
\begin{center}
\includegraphics[width=1\linewidth]{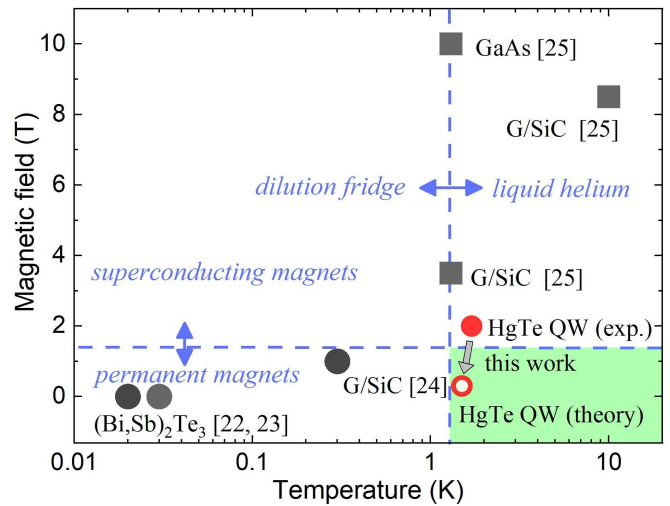}
\end{center}
\caption{Performances of several quantum Hall effect standards presented as a function of magnetic field and temperature.
Squares: demonstrated QHE standards (relative precision better than 10$^{-9}$) based on graphene and GaAs/GaAlAs quantum wells.
Circles: proof of concepts (relative precision in the range of 10$^{-4}$ - 10$^{-6}$) of QHE standards based either on graphene
with permanent magnet or ferromagnetic materials. Red solid and open circles: the experimental result and theoretical
predictions of this work. The green region is the only one accessible for both permanent magnets and liquid helium systems.}
\label{Fig1}
\end{figure}

In Fig.~\ref{Fig1} we compare performances of several quantum Hall effect (QHE) resistance standards presented as a function of magnetic field and temperature. Squares correspond to QHE standards that reached the relative precision equal or better than 10$^{-9}$ based on graphene and GaAs/GaAlAs quantum wells. Circles correspond to QHE resistance standards that reached a proof of concepts stage (relative precision below than 10$^{-9}$). They are based either on graphene with permanent magnets or ferromagnetic materials. The blue dotted lines mark the limit of magnetic field and temperature related technological barriers: superconducting magnets and dilution refrigerators correspondingly. For the dissemination of the resistance standards in wider markets, new standards should be developed that can operate at magnetic fields allowing the usage of permanent magnets and at temperatures allowing the use of liquid helium coolers. The green region in Fig.~\ref{Fig1}, is the only one accessible for both permanent magnets~\cite{Vaimann2013a} and liquid helium systems~\cite{Richardson2018a}.

As can be seen in Fig.~\ref{Fig1}, all currently used resistance standards operate at stringent experimental conditions: high magnetic fields \emph{B} (requiring the use of superconducting coils) or extremely low temperatures \emph{T} (requiring the use of dilution refrigerators), see Fig.~\ref{Fig1}. For example, typically, \emph{B} = 10~T and \emph{T} = 1.5~K for the
GaAs-based standards. Recently, using the anomalous QHE resistance~\cite{Chang2013a} of a ferromagnetic topological insulator,
feasibility of QHE resistance standards without applying  magnetic field (\emph{B} = 0~T) was demonstrated~\cite{Bestwick2015a, Martin2018a}. Additionally, QHE was observed in epitaxial graphene with $\approx$ 1~T permanent magnets~\cite{Parmentier2016a}.
However, both experiments required extremely low temperatures of $\approx$ 300 mK, provided only by dilution refrigerators. The less stringent QHE standard metrology conditions (\emph{B}~$\approx$ 3.5~T and \emph{T} $\approx$ 1.3~K) were recently achieved in epitaxial graphene~\cite{Ribeiro-Palau2015a}. This was enabled by the very large  Landau levels (LLs) splitting related to the hosting of massless Dirac fermions by the specific graphene band structure~\cite{Lafont2015a, Novoselov2007a}.
Unfortunately, suitable metrological devices were obtained only by using graphene layers deposited on SiC substrates (G/SiC), with the mobility values restricted to 10$^{4}$~cm$^{2}$V$^{-1}$ s$^{-1}$, which limited metrological applications to magnetic fields higher than 3.5~T. Furthermore, because the G/SiC surface is usually covered by resists~\cite{LaraAvila2011a}, the device properties degrade with time~\cite{Yang2016a}. Therefore, the quest for better standards is still ongoing.

The main motivation of this work was to overcome the graphene material challenges by using a two-dimensional material that mimics the graphene-like band structure but also shows higher carrier mobility. Another motivation was in research for a semiconductor material with well-established epitaxial fabrication and device processing techniques providing, with a high wafer yield, cost-efficient QHE resistance standard devices with identical geometry, contact quality and a long sensor lifetime.

Due to the recent progress in MBE), (Cd,Hg)Te-based QWs with a low carrier concentration and a very high carrier mobility can be
grown, allowing for the observation of quantum limit $\nu$~= 1 conditions in magnetic fields below 1 T~\cite{Kozlov2015a}.
Additionally, QHE in HgTe QWs with graphene-like band structures was reported~\cite{Buttner2011a}, and there structures were
recently observed even at liquid nitrogen temperatures~\cite{Kozlov2014a, Khouri2016a}. These findings show that HgTe/(Cd,Hg)Te
QWs are promising candidates for use in QHE metrology.
In this work, we demonstrate that HgTe/(Cd,Hg)Te QWs with the thicknesses close to the \emph{d$_{c}$} and trivial band
ordering, eliminating  coexistence of conducting topological helical edge states together with QHE chiral states, allows
reaching very favorable cryomagnetic conditions for the QHE resistance standards operation: low enough magnetic fields enabling
use of permanent magnets (B~$\leq$ 1.4~T) simultaneously with liquid helium coolers (teperatures of T~$\geq$ 1.3~K) – see the
green region and the circles in Fig.~\ref{Fig1}.
\section{RESULTS AND DISCUSSION} \label{result}
\subsection{Investigated structures} \label{InvSTR}
Quantum structures were grown by MBE on GaAs (013) substrates~\cite{Dvoretsky2010a}. Three samples \emph{S1}, \emph{S2} and
\emph{S3}, containing HgTe QWs with the thicknesses and barrier compositions of 7.1~nm, 8.0~nm, HgTe/Cd$_{0.62}$Hg$_{0.38}$Te, and 6.5~nm, HgTe/Cd$_{0.65}$Hg$_{0.35}$Te, respectively, have been studied. Measurements were performed on lithographically defined Hall bars with the dimensions of \emph{L} $\times$ \emph{W} = 80 $\times$ 20~\textmu m$^{2}$, and \emph{L} $\times$
\emph{W} = 40 $\times$ 20~\textmu m$^{2}$ (samples \emph{S1} and \emph{S2}, respectively) and \emph{L} $\times$ \emph{W} = 650~$\times$ 50~\textmu m$^{2}$ (sample \emph{S3}). Additionally, samples \emph{S2} and \emph{S3} had a top gate electrode allowing
for carrier density control ~\cite{Majewicz2014a}.

\begin{figure}%[H]
\includegraphics[width=1\linewidth]{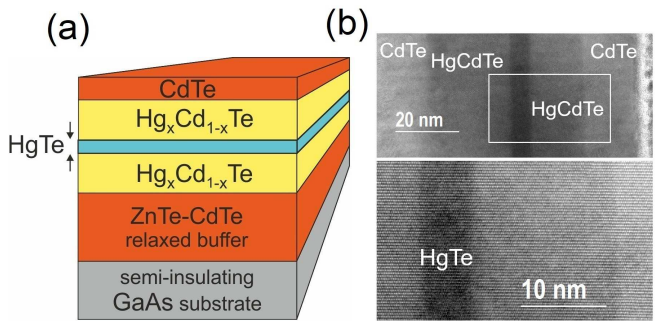}
\caption{(a) Schematic view of the layer sequence for MBE-grown HgTe/(Hg,Cd)Te QW structures; (b) High-resolution electron
microscopy images of the structure cross-section.}
\label{Fig2}
\end{figure}

Figure.~\ref{Fig2}a and .~\ref{Fig2}b show the layer sequence scheme and high-resolution electron microscopy (HREM) images of the cross section of sample \emph{S1}. A perfect alignment of the successive atomic layers with the well-defined HgTe QW can be seen in Fig.~\ref{Fig2}b. Such high growth quality is crucial for obtaining the appropriate carrier mobility of HgTe QWs for
QHE standards and can be reached only by the MBE growth technique.
\subsection{Contribution of the TI states in QHE measurements} \label{ContrTI}

\begin{figure}%[H]
\begin{center}
\includegraphics[width=0.85\linewidth]{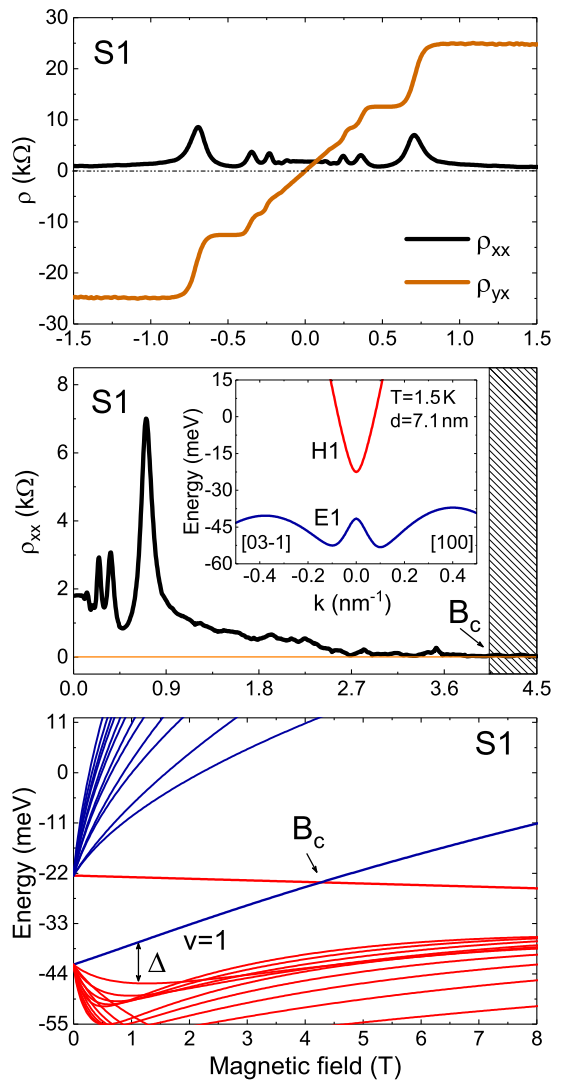}
\end{center}
\caption{(a) Magnetic field dependence of the Hall resistance $\rho_{yx}$ and longitudinal resistance $\rho_{xx}$ for sample
\emph{S1} with the inverted band structure. (b) $\rho_{xx}$ (\emph{B}) for sample \emph{S1} in a wider magnetic field range.
The inset shows calculated energy dispersion (c)  Landau levels as a function of the magnetic field for sample \emph{S1}.}
\label{Fig3}
\end{figure}

Most of the magnetotransport experiments were performed using an 8~T superconducting magnet system and standard lock-in
measurement techniques. High-precision resistance measurements were performed using an HP3458A multimeter. The longitudinal and
transverse resistances $\rho_{xx}$(\emph{B}) and $\rho_{yx}$(\emph{B}) for sample \emph{S1} are shown in Fig.~\ref{Fig3}a. The
data correspond to \emph{T} = 1.7~K, the hole concentration of $3.4\cdot10^{10}$~cm$^{-2}$ and mobility of
14.9~m$^{2}$V$^{-1}$s$^{-1}$. A well-developed QHE plateau is visible already at B $\approx$ 0.7~T. However, the $\rho_{yx}$
quantization is not exact, and $\rho_{xx}$ does not go to zero.

\begin{figure}%[H]
\begin{center}
\includegraphics[width=0.75\linewidth]{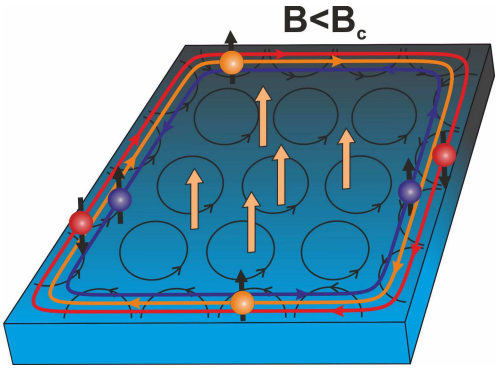}
\end{center}
\caption{A schematic drawing of the coexistence of helical quantum spin Hall edge states (blue and red lines) with a
conventional chiral quantum Hall state (orange line).}
\label{Fig4}
\end{figure}

In the search for the origin of this parasitic $\rho_{xx}$ conductivity, we consider the possibility of conduction through
topological insulator 1D edge channels, as shown schematically in Fig. ~\ref{Fig4}.

\begin{figure}%[H]
\begin{center}
\includegraphics[width=0.85\linewidth]{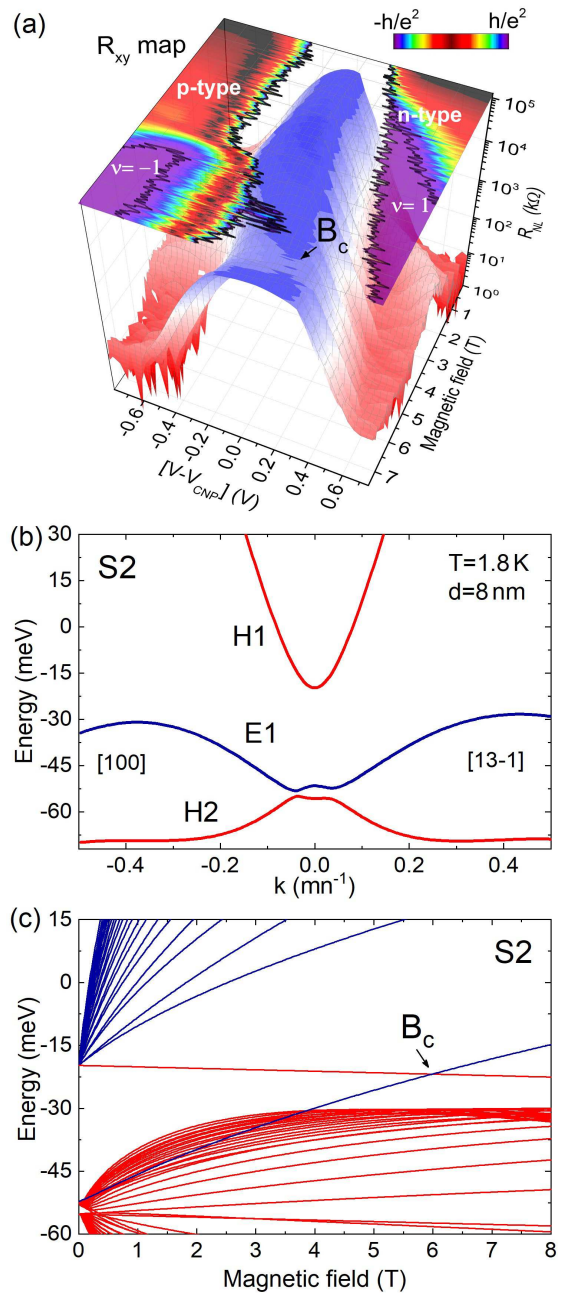}
\end{center}
\caption{a) Dependence of the nonlocal resistance on the gate voltage and magnetic field for sample \emph{S2}
(three-dimensional plot) and Hall resistance map (two-dimensional surface) showing regions of QHE for electrons ($\nu$ = 1),
holes ($\nu$ = -1). (b) Calculated energy dispersion and (c) Landau level positions as a function of the magnetic field for
sample \emph{S2}.}
\label{Fig5}
\end{figure}

We calculate the band structure and LLs fan charts for samples \emph{S1} - \emph{S3} using the eight-band $k \cdot p$
Hamiltonian~\cite{Krishtopenko2016c}, including the $\Gamma_{6}$, $\Gamma_{8}$ and $\Gamma_{7}$ bands of bulk materials and
strain effects due to lattice constant mismatch, with the parameters confirmed by our earlier
investigations~\cite{Kadykov2018a, Marcinkiewicz2017}. According to the calculation results presented in Fig.~\ref{Fig3},
sample \emph{S1} has an inverted band structure up to the critical magnetic field $B_c$~$\simeq$~4~T, at which a crossing of
the highest LL from the valence band with the lowest LL from the conduction band occurs~\cite{Konig2007a, Konig2008a,
Brune2011a, Zholudev2012a, Chen2012a}. Interestingly, $\rho_{xx}$ tends to be zero precisely when the magnetic field approaches
$B_c$ $\simeq$ 4~T (see Fig.~\ref{Fig3}b). This suggests that the existence of the inverted band structure and the related
topological gap are detrimental to good quantization. It is possible that the combination of both counter-propagating
topological edge states and disorder promotes backscattering beyond the expected topological gap~\cite{Ma2015a}.
To further explore conduction through the edge channels, we have investigated sample \emph{S2}, which also has an inverted band structure similar to sample \emph{S1}, but with a slightly larger negative energy gap. This sample was equipped with a top gate controlling the carrier type and density. Well-developed quantum resistance plateaus $\rho_{xx} = \pm h / e^2$ were observed at magnetic fields starting from above $\simeq$ 2~T for electrons and above $\simeq$ 5.5~T for the holes, as illustrated by a 2D map in Fig.~\ref{Fig5}a. The edge state conduction is revealed by the results of the nonlocal resistance measurements~\cite{Grabecki2013a} shown in the 3D plot in Fig.~\ref{Fig5}a. The value of $R_{NL}$ is plotted on the logarithmic scale to better visualize the residual edge conducting channels extending to the QHE regions. As shown in Fig.~\ref{Fig5} the resistance becomes very large when the Fermi level lies in the topological gap and decreases for magnetic fields higher than the critical value ($B_c$ $\approx$ 6~T for sample \emph{S2}). As can be seen, in spite of the nonlocal resistance has its maximum for the Fermi level close to the charge neutrality point, the nonlocality still persists far from the gap, in the QHE regions shown by the 2D map plotted above the 3D nonlocal resistance graph.

\begin{figure*}%[h!]
\includegraphics[width=1\linewidth]{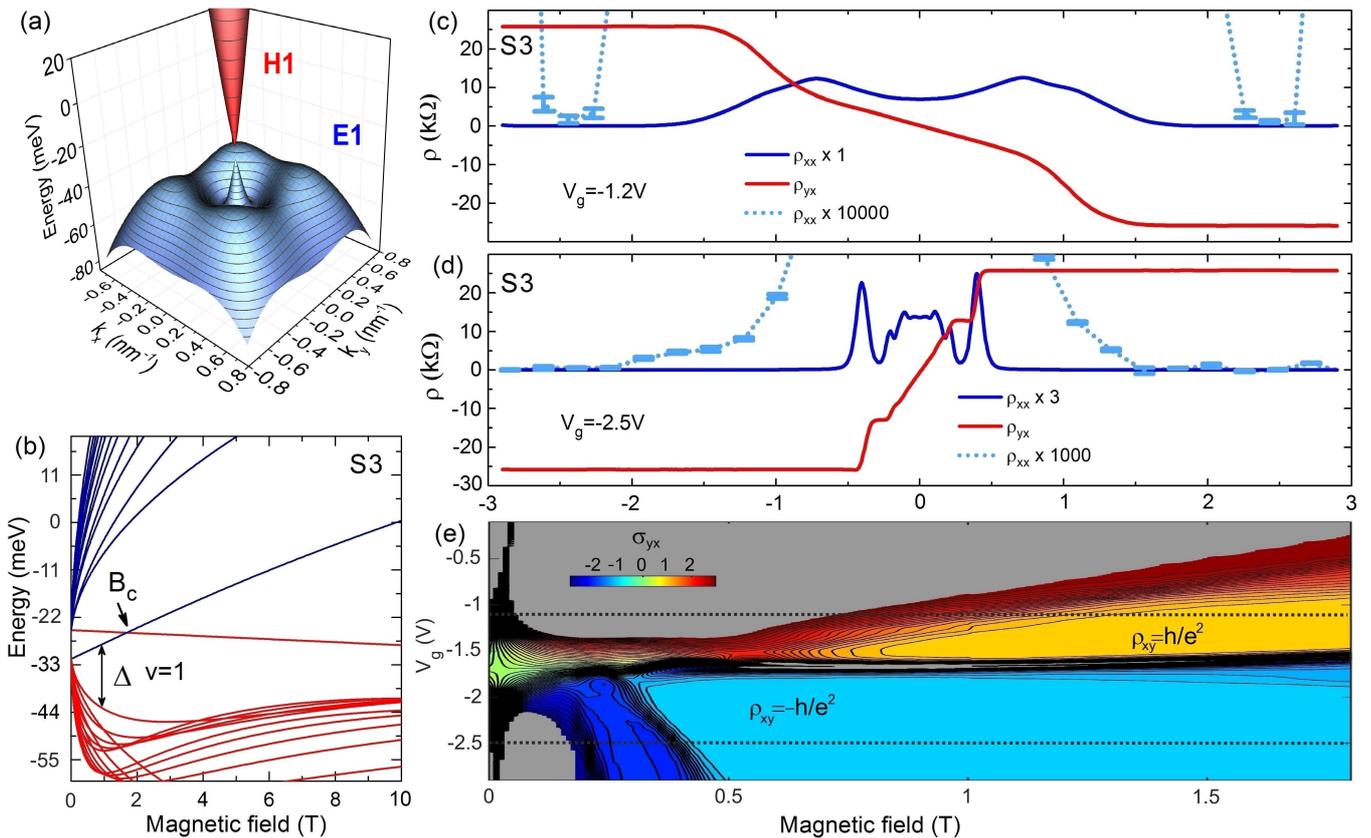}
\caption{(a, b) Calculated energy dispersion and Landau levels as a function of the magnetic field for sample \emph{S3}.
Quantum Hall effect in sample \emph{S3} at 1.7 K for two gate voltages: (c) corresponds to Fermi level in the conduction band,
(d) corresponds to Fermi level in the valence band. Light-blue dotted lines represent magnified $\rho_{xx}$(\emph{B}) at the
minima. (e)  Hall resistivity as a function of magnetic field for different Fermi level positions controlled by the gate
voltage. QHE plateaus at $\nu$ = 1 correspond to yellow and cyan areas for electrons and holes, respectively.}
\label{Fig6} \end{figure*}

\subsection{6.5 nm QW in magnetic field} \label{QWMF}
Based on results presented above, we conclude that HgTe QWs with the smallest possible $B_c$ should be engineered to decrease or avoid topological edge states perturbing the precision of the QHE. Figure~\ref{Fig6} shows the band structure of sample \emph{S3}. This sample has the smallest gap and the smallest critical magnetic field of the three investigated samples, $B_c$~$\approx$ 1.7~T. The magnetoresistance data for sample \emph{S3} are shown in Figure~\ref{Fig6}. On the electron side (V$_g$~= -1.2~V, $\mu$~= 5~m$^2$V$^{-1}$s$^{-1}$, panel c), the $\nu$~= 1 plateau is observed above 1.5~T. The longitudinal resistance $\rho_{xx}$ is less than 0.1~$\Omega$ in \emph{B} = 2.5~T, and $R_H$ = 25,807 $\pm$ 7~$\Omega$, yielding the relative accuracy in the range of $\Delta R_H/R_K~\simeq (-2~\pm 3)~\times 10^{-4}$.

On the hole side (V$_g$~= -2.5~V, $\mu$~= 20~m$^2$V$^{-1}$s$^{-1}$, panel d), the QHE is observed at a very low magnetic field \emph{B}~$\simeq$ 0.5~T. However, there is a small residual contribution $\rho_{xx}$~$\simeq$ 20~$\Omega$ at this field, which we attribute to the counter-propagating topological edge states. Indeed, this residual resistivity decreases with magnetic field and is finally strongly suppressed, dropping to less than 0.1~$\Omega$ around \emph{B} $\simeq$ 2~T. Band structure calculations show that this field corresponds to the crossing of the conduction and valence band levels as well, signifying the transition from the inverted to the trivial band structure (\emph{B}~$\simeq$ 1.7~T, Fig.~\ref{Fig6}b).

Figure~\ref{Fig6}e illustrate that QHE at $\nu$~= 1 quantum condition for p-type HgTe QWs with a width close to the $d_c$ take place at lower magnetic field than n-type ones. This fact and wide magnetic field region of QHE plateau for p-type we tentatively attribute to the presence of the valence band side maxima playing the role of a charge reservoir. Such a “reservoir” was also observed in G/SiC. It was proven that its existence is favourable for the formation of wide QHE plateaus~\cite{Kopylov2010a, Yang2016a}.

Summing all the $R_H$ values collected between 1.8~<~B~< 3~T at $I_{dc}~= \pm1$~\textmu A, we obtain $\Delta R_H/R_K~\simeq (3 \pm
30)~\times 10^{-5}$. To further increase the overall precision, it would be mandatory to f{abricate the QWs with non-inverted
band ordering and the highest possible Fermi velocity.
\subsection{Strain engineering for metrology applications} \label{MetrolApp}

The results presented above show that good quantization precision at fields close to 0.5~T can be obtained by material
engineering (slightly narrower QWs with the non-inverted band structure, $B_c \simeq $~0~T). Further improvements in the QHE
metrological conditions, i.e., lowering the operating magnetic field and increasing the temperature can be obtained in the QWs
with higher Fermi velocity $\nu_F$. This velocity is determined by the slope of the linear energy dispersion. Higher slope and
higher $\nu_F$, lead to increase of the energy spacings between the LLs. Recently, it has been experimentally demonstrated that
the use of CdTe-Cd$_{0.5}$Zn$_{0.5}$Te strained-layer superlattices on GaAs as virtual substrates with an adjustable lattice
constant allows effective control of the strain in the HgTe QWs from tensile ($\varepsilon$~< 0, $\varepsilon \approx$
–0.32~\% for CdTe buffer) to compressive ($\varepsilon$~>~0, up to +1.40~\%).~\cite{Leubner2016a} Following this idea, we have
performed calculations of the critical width $d_c$ and the Fermi velocity $\nu_F$ for the QWs with 2D massless Dirac fermions
($B_c$~= 0~T) at different strains and at temperatures up to 100~K. The obtained results are shown in Figure ~\ref{Fig7} for
the QWs grown along the (001) direction.~\cite{Krishtopenko2016c} It is worth noting that the results for the (013) QWs are
qualitatively the same. It can be seen in Fig.~\ref{Fig7} that the compressive strain indeed leads to an increase in the
$\nu_F$ of the 2D massless Dirac fermions.

\begin{figure}%[H]
\begin{center}
\includegraphics[width=1\linewidth]{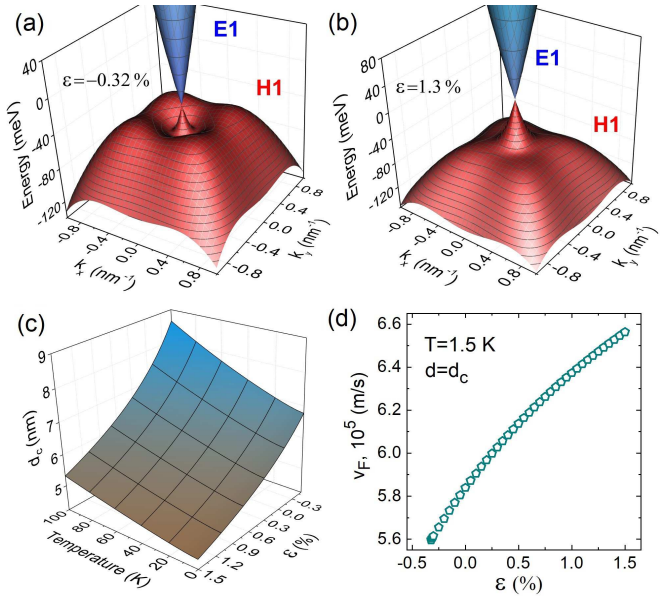}
\end{center}
\caption{((a, b)  Calculated energy dispersions for two HgTe QWs hosting massless Dirac fermions for $d_c$ = 6.4~nm, strain $\varepsilon$ = -0.32~\% (CdTe buffer) and \emph{d} = 4.6~nm, $\varepsilon$ = 1.3~ \%,~\cite{Grabecki2013a} respectively. (c) Critical QW width $d_c$ as a function of strain and temperature. (d) Band velocity $\nu_F$ of massless Dirac fermions vs.
strain.}
\label{Fig7} \end{figure}

To illustrate the advantages of HgTe QWs as the QHE resistance standards, in Fig.~\ref{Fig8}, we compare the energy spacing between the LLs ($\Delta$) at the quantum limit conditions ($\nu$ = 1) for G/SiC, GaAs/(Al,Ga)As QWs and the investigated HgTe/(Cd,Hg)Te QWs. It is conventionally assumed that thermal excitations cease to be relevant for metrological applications
for $\Delta \geq$ 100~$k_BT$, corresponding to $\Delta \geq$ 13~meV at \emph{T}~= 1.5~K. The second condition is $\mu B$ > 1, where $\mu$ is the carrier mobility that determines the degree of the LL energy broadening. Both conditions, $\mu B$~> 1 and $\Delta \geq$ 100~kBT, merge together, defining the metrological regions marked by arrows in Figure~\ref{Fig8}.
\begin{figure}%[H]
\begin{center} \includegraphics[width=1\linewidth]{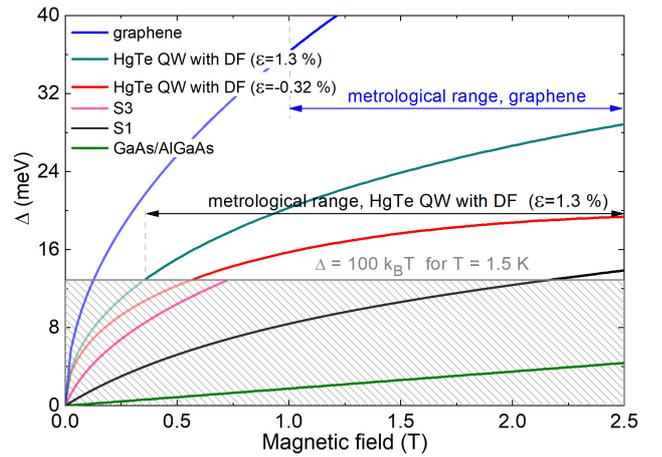}
\end{center}
\caption{Energy gap $\Delta$ between LLs at $\nu$= 1 as a function of the magnetic field for these two QWs (solid and dashed lines, respectively) as well as for samples \emph{S1} (diamonds) and \emph{S3} (stars); data for graphene (dotted line) and GaAs/AlGaAs (dashed-dot line) are also shown. The grey horizontal line is the lower boundary of the metrological range $\Delta \geq$ 100$k_BT$ at \emph{T} = 1.5 K. Vertical dashed lines indicate metrologically useful ranges, where conditions $\Delta \geq$ 100$k_BT$ and $\mu B$ > 1 are both fulfilled. Mobility values for graphene (G/SiC) are estimated from
Ref.~\cite{Ribeiro-Palau2015a}}
\label{Fig8} \end{figure}

The second, the magnetic field condition, is a limiting factor for G/SiC for which the $\mu \approx$~1~m$^2$V$^{-1}$s$^{-1}$ and $\mu B$~>~1 conditions areis fulfilled only for a magnetic field \emph{B}~ >~1~T. In HgTe QW samples, (mobility ~10~m$^2$V$^{-1}$s$^{-1}$) this condition is fulfilled already at $B~\geq$~0.1~T. One can see that the metrological range for HgTe QWs samples may be obtained in relatively lower magnetic fields.

The optimum \emph{T}  $\approx$ 1.5~K metrological conditions for HgTe QWs are predicted with \emph{d}~= 6.4~nm ($\varepsilon$~= -0.32~\%) and \emph{d}~= 4.6~nm ($\varepsilon$ = 1.3~\%) for \emph{B}  $\approx$ 0.3~T and $B \approx$~0.5~T respectively.
These QWs are hosting Dirac fermions with normal band ordering, ensuring maximum landau level splittings at absence of topological edge conducting channels. In Fig.~\ref{Fig7}c) we show results of calculations of  the critical thickness of the
HgTe QWs as a function of the strain and temperature.
 \section{CONCLUSION} \label{conclusion}
We have investigated the HgTe/(Cd,Hg)Te~QWs in view of their application for new competitive QHE standards. We
have shown that the coexistence of helical quantum spin Hall states and conventional chiral quantum Hall states in the
topologically non-trivial region (\emph{d}~>~$d_c$, $B~< B_c$) may restrict the accuracy of the Hall resistance quantization.
At the same time, our results demonstrate that for the non-inverted QWs close to the critical width $d_c$, at which Dirac fermions
control the charge transport, the QHE metrological conditions can be reached for holes (p-type samples) in very favorable cryomagnetic conditions: low magnetic fields (going even below 0.5~T) and at relatively high temperatures (going even above 1.5~K). These conditions are substantially more favourable than those required for the use of current QHE standards. Our experiments and realistic band structure calculations indicate a possibility for further improvements of the QHE metrological conditions in compressively strained HgTe/(Cd,Hg)Te~QWs, for which the Fermi velocity is even higher. This opens the door for the development of QHE metrological devices operating with permanent magnets (without superconducting coils) and at liquid helium temperatures (without dilution refrigerators). Accordingly, MBE-grown (Hg,Cd)Te-based~QWs with Dirac fermions are semiconductor nanostructures with prospects for a large scale fabrication  and a wide use of QHE resistance standards devices by academic and industrial institutions.
\begin{acknowledgments}
The research in Poland was partially supported by the Foundation for Polish Science through the IRA Program financed by EU
within SG~OP Program and No.~TEAM/2016-3/25, by the National Science Centre, Poland, decisions No.~DEC-2011/02/A/ST3/00125,
2013/10/M/ST3/00705, UMO-2015/17/N/ST3/02314, 2016/22/E/ST7/00526 and UMO-2017/25/N/ST3/00408. The authors also acknowledge
Russian Foundation for Basic Research (Grants 15-52-16008 and 16-02-00672), ARPE Terasens project and Terahertz Platform both
provided by Occitanie Region. The theoretical calculations for the compressively strained QWs were performed in the framework
of project 16-12-10317, supported by the Russian Science Foundation. S.~S.~Krishtopenko and A.~Kadykov acknowledge the Russian
Ministry of Education and Science (MK-1136.2017.2 and SP-5051.2018.5). The authors gratefully thank Z.~D.~Kvon from the
Institute of Semiconductor Physics (Siberian Branch, Russian Academy of Sciences) for stimulating and fruitful discussions as
well as for providing the processed gated Hall bar samples.
\end{acknowledgments}
\appendix
\section{APPENDIX} \label{appendix}
Figure~\ref{Fig5} in the main text presents the behaviour of non-local resistance vs magnetic field and gate voltage. In
conventional semiconductors, nonlocal resistance shows a well-pronounced zero, because the resistance decays exponentially with
distance L between the two Hall arms according to $R_{NL} = R_{sq}$~exp$(-\pi L/W)$, where $R_{sq}$ is the sheet resistance,
and \emph{W} is the width of the Hall bar~\cite{VanderPauw1958a}. In the experiment, $R_{NL}$ was defined as the ratio between
the measured voltage on 3-4 contacts and applied current 10~nA on 1-2 contacts (see Figure~\ref{Fig9}).

\begin{figure}[h!]
\begin{center} \includegraphics[width=0.75\linewidth]{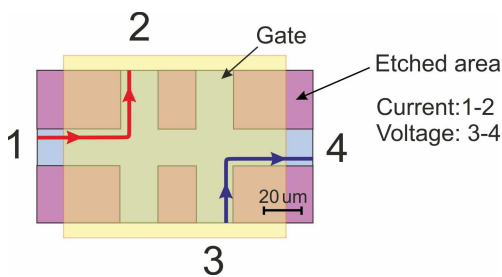}
\end{center}
\caption{Schematic configurations of nonlocal resistance measurements for sample \emph{S2}. Current was applied to contacts
1-2, voltage was measured on contacts 3-4}
\label{Fig9} \end{figure}

It should be stressed that for the samples with the band structure close to the Dirac band structure, the holes provide more
favourable conditions for the QHE resistance standards, namely, lower magnetic fields and larger plateaus. The results
presented in Figure~\ref{Fig6}(c-e) show an asymmetry between the QHE results for holes (panel d) and electrons (panel c). It
may be observed that for holes, the QHE plateaus are wider and start at lower magnetic fields. We attribute this characteristic
to the specific details of the HgTe valence band structure. The valence band side maxima play the role of a charge reservoir.
Such a “reservoir” was also observed in G/SiC. It was proven that its existence is favourable for the formation of wide QHE
plateaus~\cite{Yang2016a, Kopylov2010a}. Thus, a tuning of the energy difference between the position of the side maxima in the
valence band and charge neutrality point (marked as $\delta_{VB}$, see Fig.~\ref{Fig10}) as a function of the strain in the
HgTe layer is a powerful approach for controlling/adjusting the QHE plateau width. Figure~\ref{Fig10} illustrates the
implementation of the compressive strain into the system, giving rise to the increase in the $\delta_{VB}$ and the
disappearance of the side maxima.
\begin{figure}[h!]
\begin{center} \includegraphics[width=1\linewidth]{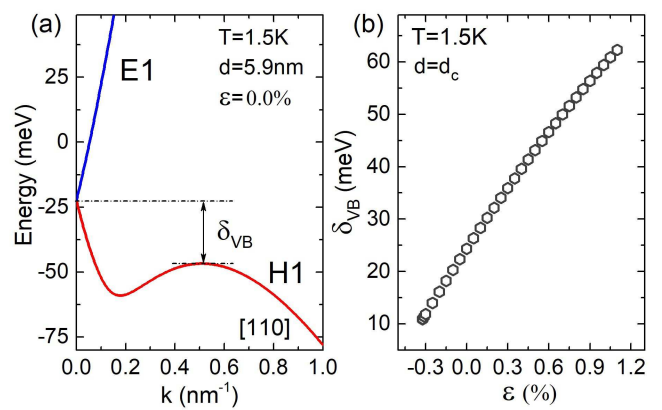}
\end{center}
\caption{a) Band structure for (001) HgTe/Cd$_{0.65}$Hg$_{0.35}$Te QWs at \emph{T} = 1.5~K. (b) Difference in energy between
the position of the side maxima in the valence band and the charge neutrality point as a function of the strain in the HgTe
layer. The lowest negative $\varepsilon$~= -0.32~\% corresponds to the QW grown on the CdTe buffer. The results for the QWs
grown along (013) are qualitatively the same}
\label{Fig10} \end{figure}

Although we have demonstrated that HgCdTe QWs have the potential to be new competitive resistance standards, further material engineering is still necessary to obtain practical metrological QHE standards. Let us discuss briefly the most important and challenging developments. The first goal of material engineering is to further decrease the longitudinal resistance. It can also be seen that this parasitic resistance depends on the Fermi level and decreases by almost an order of magnitude, with the gate voltage changing from -2.2~V to 2.5~V. This indicates that in the metrological samples, an even better accuracy of $\Delta R_H/R_H$ can be obtained by playing with the relative positions of the central and side maxima~\cite{Krishtopenko2018a} as well as the doping level. Additionally, tuning of the carrier density with gates may lead to undesired current leakages, generating errors in the resistances and/or noise. Therefore, once the band structure and the optimal hole density are established, the samples for the final metrological devices should be grown and processed without the gate.

%%%%%%%%% References %%%%%%%%%

%\bibliographystyle{apsrev4-1}
%\bibliography{mybib}
%

\end{document}